# An Extended Phase Graph-based framework for DANTE-SPACE simulations including physiological, temporal, and spatial variations

Matthijs H.S. de Buck[1,*], Peter Jezzard[1], and Aaron T. Hess[1]

[1]Wellcome Centre for Integrative Neuroimaging, FMRIB Division, Nuffield Department of Clinical Neurosciences, University of Oxford, Oxford, United Kingdom

Word Count: 4767

Figures: 10

**Correspondence address**:

Aaron T. Hess, PhD

FMRIB Division, Wellcome Centre for Integrative Neuroimaging

University of Oxford

John Radcliffe Hospital

Oxford, OX3 9DU

United Kingdom

Email: aaron.hess@ndcn.ox.ac.uk

* Current address: Spinoza Centre for Neuroimaging, Meibergdreef 75, Amsterdam, The Netherlands

# Abstract

**Purpose:** The DANTE-SPACE sequence facilitates three-dimensional intracranial vessel wall imaging with simultaneous suppression of blood and cerebrospinal fluid (CSF). However, the achieved image contrast depends closely on the selected sequence parameters, and the clinical use of the sequence is limited *in vivo* by observed signal variations in the vessel wall, CSF, and blood. This paper introduces a comprehensive DANTE-SPACE simulation framework, with the aim of providing a better understanding of the underlying contrast mechanisms and facilitating improved parameter selection and contrast optimization.

**Methods:** An Extended Phase Graph (EPG) formalism was developed for efficient spin ensemble simulation of the DANTE-SPACE sequence. Physiological processes such as pulsatile flow velocity variation, varying flow directions, intravoxel dephasing, diffusion, and $B_1^+$ effects were included in the framework to represent the mechanisms behind the achieved signal levels accurately.

**Results:** Intravoxel velocity averaging improved temporal stability and robustness against small velocity changes. Time-varying pulsatile velocity variation affected CSF simulations, introducing periods of near-zero velocity and partial rephasing. Inclusion of diffusion effects was found to substantially reduce the CSF signal. Blood flow trajectory variations had minor effects, but $B_1^+$ differences along the trajectory reduced DANTE efficiency in low-B1+ areas. Introducing low-velocity pulsatility of both CSF and vessel wall helped explain the *in vivo* observed signal heterogeneity in both tissue types.

**Conclusion:** The presented simulation framework facilitates a more comprehensive optimization of DANTE-SPACE sequence parameters. Furthermore, the simulation framework helps to explain observed contrasts in acquired data.

# 1. Introduction

The DANTE-SPACE sequence facilitates three-dimensional intracranial vessel wall imaging by simultaneously suppressing the blood inside the vessel and the surrounding cerebrospinal fluid (CSF). DANTE-SPACE consists of a DANTE[1,2] (delay alternating with nutation for tailored excitation) module for suppressing moving spins followed by a variable-flip-angle turbo-spin-echo SPACE readout[3]. It has been used at both 3T[2,4–15] and 7T[16–19], and with SPACE parameter configurations resulting in T1-weighted[4–11,15,19], T2-weighted[2,13,16–18], and proton density-weighted[2,10,12,15] contrasts. Across different implementations, many different protocol settings have been used, such as DANTE pulse trains ranging from 64 pulses[2] to 300 pulses[2,14–16] and DANTE flip angles ranging from 8°[5,6,8,15] to 14°[19]. The choice of these parameters directly affects the resulting blood, CSF, and vessel wall (VW) signals. Various authors have used DANTE calculations[1,12], Bloch simulations[2,4,8,13–16,19], or direct comparison of acquisitions using multiple parameter combinations[15] to select parameters aiming to achieve the highest contrast between the vessel wall, blood and CSF.

The validity and accuracy of such approaches rely on design approximations and assumptions. Li et al.[1] derived a T1-decay model to approximate signal decay in moving spins during playout of the DANTE preparation. This approach is both intuitive and computationally efficient. However, it assumes complete velocity-independent spoiling of the transverse magnetization for moving spins. This can be inaccurate for pulsating or very slowly moving spins. Furthermore, this model only accounts for the effects of DANTE, without taking into account the effects of readout pulses and multiple repetitions. More accurate models can be achieved using Bloch simulations (either including[4,8] or excluding[14–16] readout effects) which directly account for flow effects in the simulations. This requires tissue-specific flow properties as input parameters. For this, CSF is often assumed to be either static[8] or moving substantially faster (≥2 cm/s[14–16]) than the values found in the literature for the CSF near the Circle of Willis (e.g., 0.37 cm/s in the 3rd ventricle[20] and up to 0.85 cm/s in the cerebral aqueduct[21]). Furthermore, other processes affecting measured signal levels, such as intravoxel dephasing and pulsatile flow velocity variation, are generally also not modelled in those simulations. At 7T another possibly non-negligible factor is the effect of transmit field variation on the obtained MRI signals and contrasts. Although this can be expected to have a limited effect near the Circle of Willis (located in the central region of high $B_1^+$ efficiency at 7T when using a transmit head coil), it can affect the magnetization of blood spins, which can travel from low-$B_1^+$ areas in the neck into higher-$B_1^+$ areas near the Circle of Willis during a single DANTE-SPACE module.

This paper introduces a comprehensive DANTE-SPACE simulation framework to account for all these effects. By including physiological processes, such as pulsatile flow velocity variation, varying flow directions, intravoxel dephasing, and diffusion, as well as $B_1^+$ effects, this framework attempts an accurate representation of the mechanisms behind the achieved signal levels (VW, CSF, and blood). This in turn can be used for accurate optimization of sequence parameters, and to obtain a further understanding of DANTE-SPACE contrast mechanisms.

Note that this paper contains a combined section which discusses both the development of the simulation framework and some initial tests of the resulting simulations. Within this section, the subsections describe all components of the simulation framework in turn, including the effects of each individual component on the resulting simulations.

## 2. Methods and Results

### 2.1. General simulation framework

All code was implemented in MATLAB R2019a. Extended phase graph (EPG) simulations[22,23] were used to efficiently simulate the magnetization evolution of ensembles of spin isochromats. This provides a more accurate representation of the signal behaviour in an acquired voxel than Bloch simulations of single isochromats, while being much more computationally efficient than averaging the results from hundreds of individual Bloch simulations to approximate spin ensemble behaviour. Code from the MRSignalSeqs toolbox (github.com/mribri999/MRSignalsSeqs) was used to propagate EPG states through periods of relaxation, gradient waveforms, RF pulse rotation, and diffusion. An additional tool was developed to simulate flow effects[3]. Simulations were performed using an Intel (Intel, Santa Clara, CA, USA) Xeon CPU E5-2680v4 running at 2.40 GHz with 14 cores and 28 logical processors.

All simulation results in this paper were obtained assuming typical relaxation properties and B1$^+$ field effects corresponding to 7T. T1 and T2 relaxation times at 7T for vessel wall tissue were based on carotid artery measurements presented by Koning et al.[24]. For blood and CSF relaxation times the same values as used by Viessmann et al.[16] (based on various original sources[25–27]) were used. The resulting values were:

- VW:    T1 = 1628 ms, T2 = 46 ms
- Blood: T1 = 2290 ms, T2 = 100 ms
- CSF:   T1 = 4019 ms, T2 = 311 ms

The contrast between tissue types was calculated from the amplitude of the point spread function of the transverse magnetization during the SPACE-readout for the various tissue types after correcting for differences in proton density ($PD_{VW} = 0.72 \times PD_{CSF} = 0.72 \times PD_{blood}$)[16].

## 2.2. Basic DANTE-SPACE simulations

All DANTE-SPACE simulations in this section were performed using the T2-weighted 7T protocol proposed by Viessmann et al.[16].

### 2.2.1. DANTE preparation module

EPG DANTE simulations were validated by comparing them to the results of Bloch equation ensemble simulations. The Bloch equation simulations themselves were first validated by comparing single isochromat simulation results to the Bloch simulation results presented by Li et al.[1] (Supporting Information Figure S1a).

For Bloch equation ensemble simulations, the average results of 1000 Bloch equation simulations over a distance corresponding to unit phase accumulation along the gradient direction were used. Supporting Information Figure S1b-c shows the results of such Bloch equation ensemble simulations and the corresponding EPG simulation results, indicating good agreement. The computation time, however, of the EPG results is two orders of magnitude shorter (0.4 s versus 57.5 s), which confirms that EPG simulations can provide accurate isochromat ensemble simulations with high computational efficiency.

### 2.2.2. SPACE readout module

To validate the SPACE-readout simulation using the EPG framework, the simulated transverse magnetization evolution in stationary vessel walls was cross-checked against the previously described target SPACE magnetization function[3,16]. Supporting Information Figure S2 shows that the prescribed signal evolution used to design the SPACE flip angle train agrees well with the simulated SPACE signal predicted by the EPG simulations. However, when including DANTE-preparation, the change in the magnetization state before the SPACE readout module reduces the level of agreement with the prescribed magnetization evolution. This reduces the resulting signal intensity by 29%, and slightly increases the width of the point-spread function of the vessel walls (by 6%).

### 2.2.3. Inclusion of vessel wall, CSF, and blood compartments

The simulation results for the combination of the EPG DANTE- and SPACE-modules are shown in Figure 1 for VW, CSF, and blood. The results show only the most basic simulations of the different tissue types, only assuming their respective relaxation times and (constant) average velocities (from literature[20,28]).

The results in Figure 1 are shown during the 2$^{nd}$ TR period for the VW and CSF simulations, but for the 1$^{st}$ TR period for the blood. This reflects the fact that while blood flows rapidly in from the upstream arteries (resulting in 'fresh' blood spins without magnetization history arriving in the readout volume in each TR), the VW and CSF remain in the same scan region throughout an acquisition. For the VW and CSF simulations, the 2$^{nd}$ TR was chosen since it was found to provide good convergence of the resulting signal within a reasonable simulation time: for the VW, the signal reduces by 24% from the first to the second TR period, after which it changes by less than 0.01% over the next 5 repetitions. For CSF, the simulated signal reduces by 25% from the first to the second TR period, with about a 1.1% change over the next 5 repetitions (while requiring 3.3× more simulation time). Based on this argument, all simulation results presented in this work are based on 2 repetitions for VW and CSF, but only a single repetition for blood.

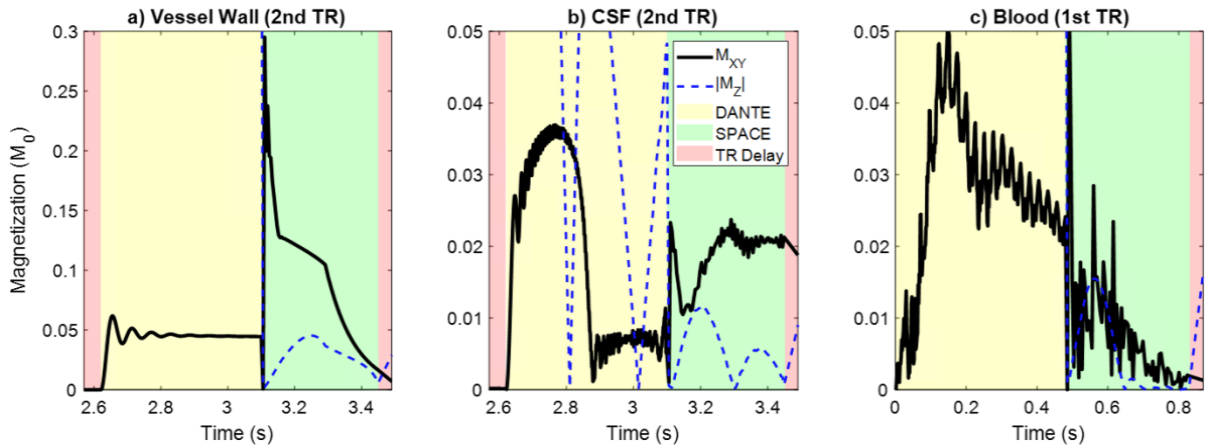

*Figure 1:* DANTE-SPACE simulations of the VW, CSF, and blood, assuming constant flow velocities (0 cm/s for VW, 0.37 cm/s for CSF, and 24 cm/s for blood) and excluding the effects of intravoxel dephasing, diffusion, pulsatility, flow trajectories, and $B_1^+$ variations. Note that a different y-axis range is shown for VW (a) than for CSF and blood (b-c).

## 2.3. Incorporation of physiological and physical effects

Next we introduce various additions to the simulations on a per-phenomenon basis. After all individual additions have been introduced, the effect is shown when all variations are included simultaneously.

### 2.3.1. Effects of intravoxel velocity variation

As is visible in Figure 1b-c, where moving spins were simulated with a specific velocity, simulations using a single velocity result in unstable predictions of transverse magnetization. However, in reality the measured MR signal in a typical voxel is generated by the transverse magnetization of spins moving with slightly varying velocities around a certain average value. To model this intravoxel velocity averaging effect in the simulations, each individual simulation was repeated 100 times using different velocity values described by a normal distribution with $\sigma = 10\%$ around the nominal velocity. This distribution and the resulting change in the magnetization evolution when accounting for a range of velocities is shown in Figure 2. The simulations with this more realistic velocity distribution demonstrate higher temporal stability of the magnetization evolution of both blood and CSF, resulting (especially for blood) in lower transverse magnetization magnitude during the SPACE readout and therefore reduced signal levels.

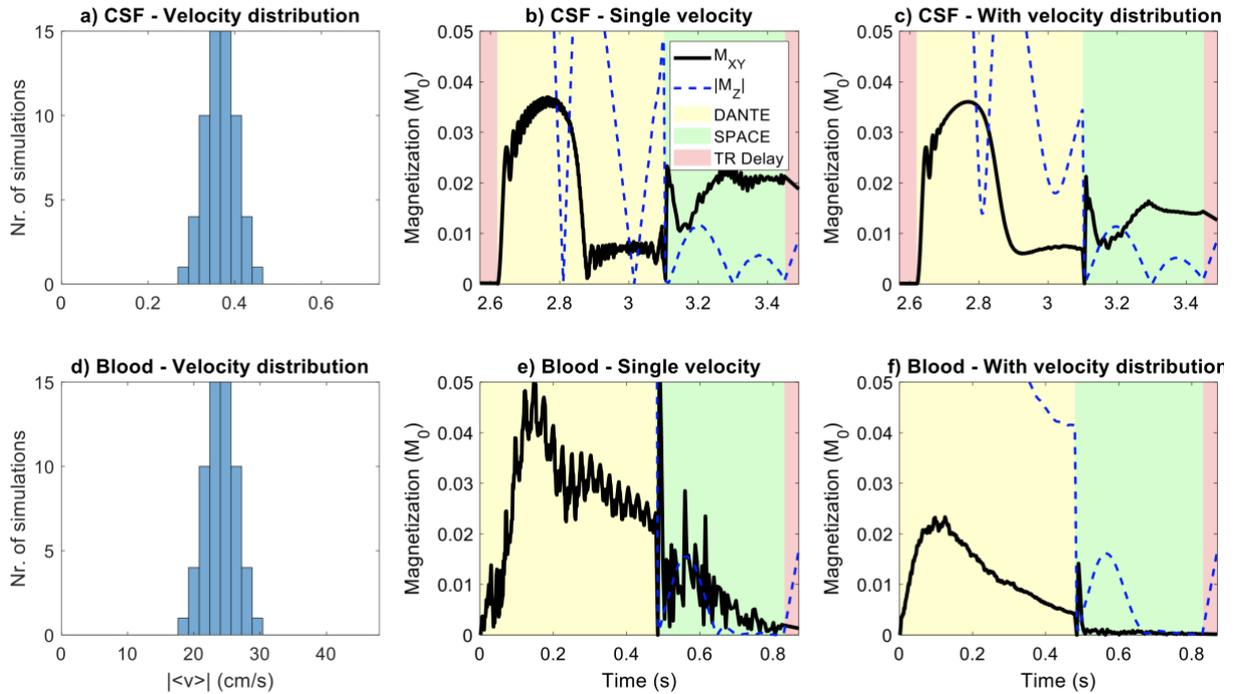

*Figure 2: Intravoxel dephasing in simulations by using the average of 100 simulations with a distribution of flow velocities. For CSF, (a) shows the distribution of velocity values (corresponding to a normal distribution with σ = 10% around the nominal velocity), (b) shows the result of a simulation using a single isolated velocity, and (c) shows the average result of the 100 simulations using the velocities from the distribution in (a). (d-f) show the corresponding results for blood.*

### 2.3.2. Effects of pulsatile motion

Pulsatile velocity variations over time were added to the simulations to represent fluid dynamics and pulsatile oscillations. Figure 3 shows the flow velocity variations that were modelled and the resulting changes in simulated signal for CSF and blood. Temporal velocity variations due to pulsatile CSF (Figure 3a) were added based on literature-sourced time-varying CSF flow measurements at the third ventricle (with an average flow velocity of 0.37 cm/s[20]). The temporal variation in blood flow velocity (Figure 3d) was modelled on the cardiac pulsatility of blood in the internal carotid arteries[28]. Both pulsation profiles assume a heart rate of 60 beats per minute. Because cardiac gating is typically not used for DANTE-SPACE acquisitions, a different random starting point was used for the pulsatile profiles in each of the 100 simulations.

Since the velocity of the blood remains relatively high throughout the cardiac cycle (with a minimum velocity of around 16.5 cm/s), the simulated magnetization of blood remains similar with and without the inclusion of pulsatility in the simulations. In contrast, CSF pulsation results in periods of both positive and negative velocity. This introduces two effects that change the signal attenuation efficiency during DANTE (which were not included in earlier versions of this simulation framework[17,18]): periods of near-zero velocity reduce the signal attenuation during DANTE, and partial rephasing can occur when the oscillations change direction. This results in a visible change in magnetization evolution when accounting for CSF pulsatility.

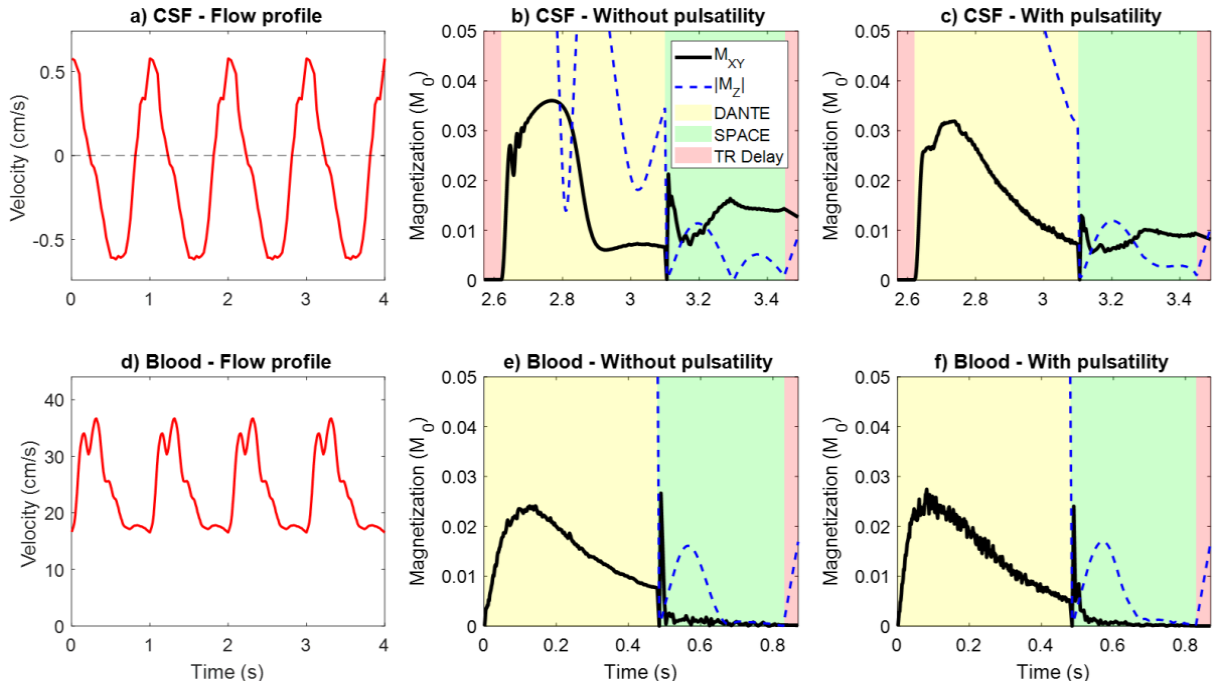

*Figure 3: Pulsatile velocity variations of (a-c) CSF and (d-f) blood, and the resulting changes in the simulated magnetization evolution for both tissue types.*

### 2.3.3. Effects of diffusion

In addition to spins moving due to flow and pulsation, the effects of diffusion were included for CSF and blood using a diffusion coefficient of $3 \times 10^{-3}$ mm$^2$/s, corresponding to free water diffusion at body temperature. For CSF, the inclusion of diffusion reduces the longitudinal magnetization of the CSF at the end of the DANTE-preparation, resulting in a 54% reduction in the measured signal during SPACE.

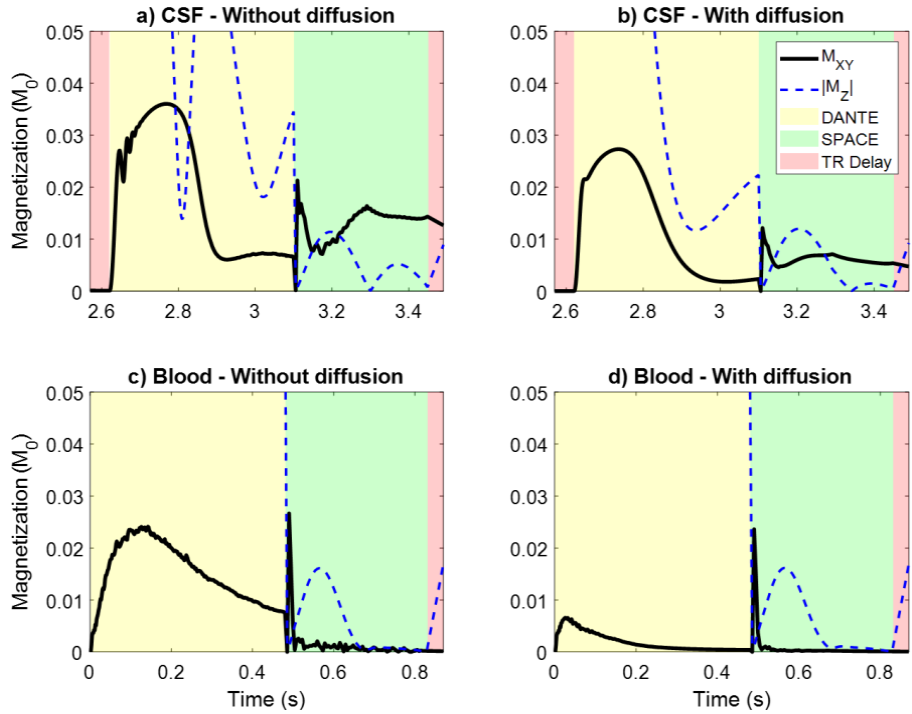

***Figure 4:*** *The effect of diffusion with a diffusion coefficient of $3 \times 10^{-3}\ mm^2/s$ on **(a-b)** CSF simulations and **(c-d)** blood simulations.*

### 2.3.4. Effects of flow trajectory

When modelling the direction of blood flow, a typical flow trajectory was incorporated into the simulations to introduce time-varying flow directions relative to the DANTE-gradient vector direction. This flow trajectory was determined from the bright-blood MPRAGE data as shown in Figure 5a. Blood was assumed to flow along this trajectory at the mean flow velocity used in simulations (24 cm/s), resulting in time-dependent variations in the flow direction as the blood moves along the vessel path. At each timepoint in the simulations, the resulting angle $\beta_{blood}$ between the flow direction and the gradient vector was used to modify the effective blood flow velocity along the gradient via multiplication by $\cos(\beta)$. For simplicity, only orientations within the coronal plane were assumed.

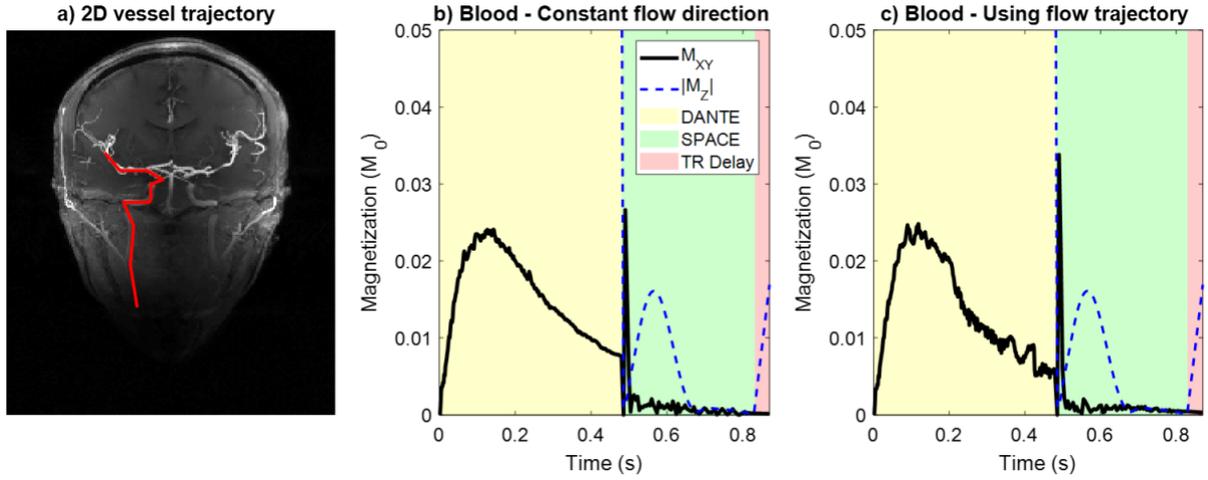

*Figure 5: The blood flow trajectory used for simulations (a), and the resulting change in the simulated magnetization (b-c).*

For CSF, directions of motion (on the relevant timescales) are known to vary throughout the brain[28]. To simplify the simulations while accounting for this variation, the angle $\beta_{CSF}$ between the CSF flow direction and the gradient vector was fixed as the average angle between a given vector (i.e., the DANTE-gradient vector) and all other possible vector directions (i.e., the possible CSF motion vectors) on a unit semi-sphere:

$$<\beta_{CSF}> = \frac{1}{V_{SS}} \int_{SS} \beta dV = 57.3°. \quad (1)$$

### 2.3.5. Incorporation of B1+ variations along the blood flow trajectory

B1+ variations in different parts of the vasculature were included in the model based on measured B1+ maps. For this, data from a previously presented database of 7T multi-channel B1+ flip angle maps in both the head and the neck were used[29]. The blood flow trajectory in Figure 5a was used to extract the B1+ values along the vessel trajectories from the B1+ maps, as shown in Figure 6a. For CSF and VW, a constant B1+ offset can be added to simulations, for example to model the effects of different pTx RF shims on the achieved CSF suppression.

The measured B1+ values were used in the simulations as a scaling factor between the nominal flip angle and the effective applied flip angle. In the neck, with typically very low B1+ at 7T, this effectively reduces the applied flip angles at the start of the DANTE trajectory. This reduces the suppression due to DANTE, which results in a slight increase in the (measured) transverse magnetization during the SPACE readout.

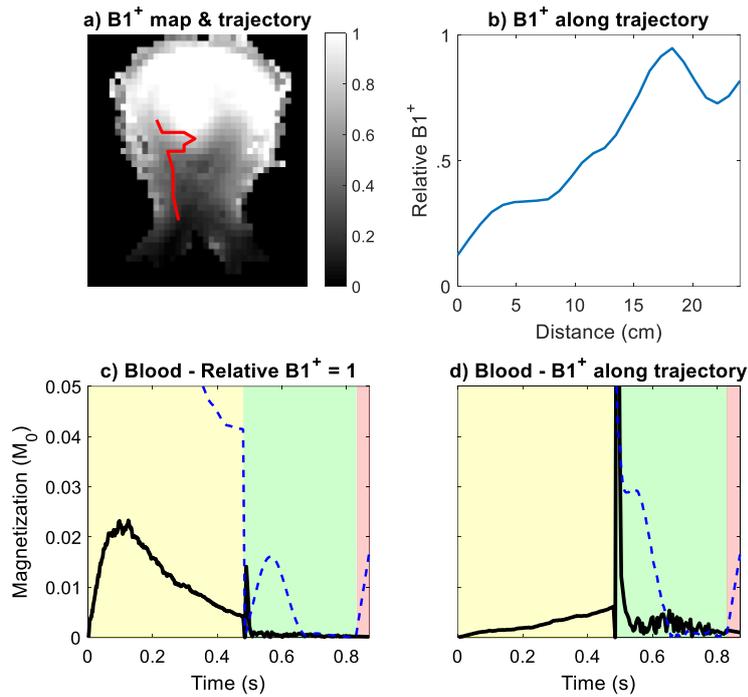

*Figure 6: $B_1^+$ variations along the 25cm blood flow trajectory. **(a)** Measured $B_1^+$ maps (in CP-mode) were used to provide values along the previously described vessel trajectories, which are shown in **(b)**. This results in a change in the simulated blood magnetization between simulations **(c)** without $B_1^+$ effects and **(d)** with $B_1^+$ effects.*

### 2.3.6. Combined variations

The previous sections discussed how intravoxel velocity distribution, pulsatile motion, diffusion, flow trajectories, and $B_1^+$ variation were implemented in the EPG simulations, while showing the isolated effect of each individual addition. Figure 7 demonstrates the effect of simultaneously including all these model enhancements. For the examples shown here (using the DANTE-SPACE protocol parameters as proposed by Viessmann et al.[16] and CP-mode $B_1^+$ maps), the combined model enhancements result in a 57% reduction in CSF signal and a 27% reduction in blood signal versus the case in which only the basic simulation from Section 2.2 is performed.

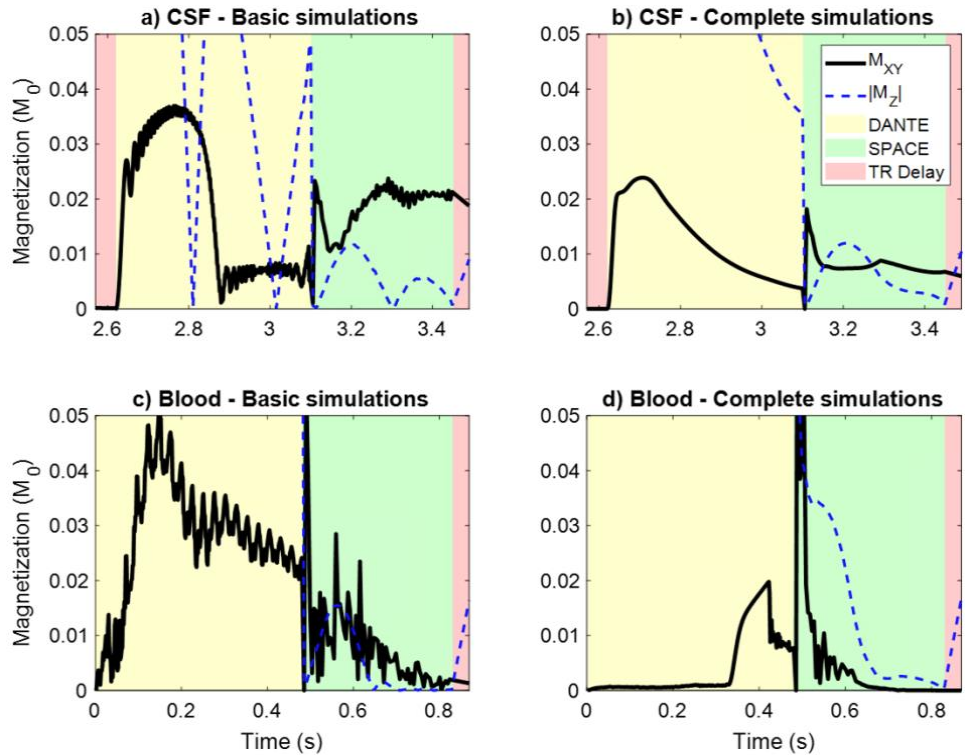

*Figure 7: The combined effect of intravoxel velocity averaging, pulsatile motion, diffusion, flow trajectories, and B1+ variation on DANTE-SPACE simulations of CSF (a-b) and blood (c-d).*

Like all CSF and blood simulation results presented here, the results in Figure 7 are shown using a small y-axis range to clearly visualize the changes in the transverse magnetization during the SPACE readout. Supporting Information Figure S3 presents the results in Figure 7 using the full y-axis (ranging from 0 to $M_0$) to also indicate the changes in the evolution of the longitudinal magnetization during the DANTE-preparation. This confirms that in the complete simulation model, the longitudinal magnetization of the blood during DANTE decreases more slowly due to the reduced efficiency of the DANTE-preparation in areas with low $B_1^+$.

## 2.4. Sensitivity to simulation parameter values

The proposed simulation model uses various tissue-specific properties to distinguish the different tissue types. This includes T1 and T2 relaxation times, mean velocities of flow or pulsation, flow directions, the amount of intravoxel dephasing, and diffusion coefficients. For VW, CSF, and blood, each of these values was carefully chosen with the aim of accurately representing the resulting signal behaviour. To assess the extent to which the simulations depend on the exact choice of tissue parameters, and to ascertain that the results are robust to small changes in the selected values (e.g., due to measurement uncertainty), this section compares simulations using variations in these tissue properties.

### 2.4.1. T1 & T2 relaxation times

As described in Section 2.1, the T1 and T2 values for VW, CSF, and blood were based on various measurement results from the literature. Each of the resulting values has a certain amount of measurement uncertainty (up to 28%[27]). Therefore, Figure 8 compares the results of simulations for all three tissue types where the T1 and T2 values are varied between the selected value ± 30%. This indicates that the results are largely independent of the T1 of blood and the T2 of both CSF and blood. The T2 dependence of the vessel wall signal is consistent with the T2-weighted SPACE readout used in the simulations (which is calculated based on vessel wall relaxation times[16]). Despite this T2-weighting, Figure 8 also shows some T1-dependence of the VW and CSF signals. This is mainly a result of the DANTE-preparation, during which a higher T1 results in increased suppression of the longitudinal magnetization as well as reduced longitudinal magnetization recovery during the TR delay. Because of its higher T1 and increased DANTE sensitivity, this T1 dependence is more pronounced in CSF. However, the resulting CSF signal remains much smaller than the VW signal for all simulated T1 values.

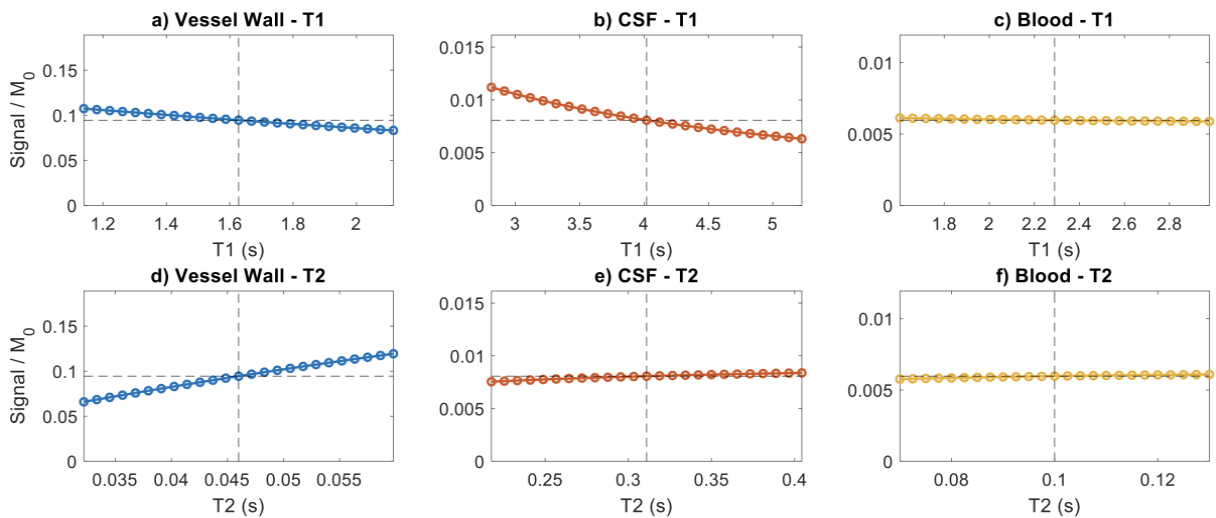

*Figure 8:* *Simulation results when using different T1 and T2 relaxation times. Results are shown for each tissue type for the standard values ± 30% (x-axes). Different y-axes are used for the different tissue types, all of which are scaled from 0 to twice the simulated signal using the standard (T1/T2) values. Dashed lines indicate the standard T1/T2 values and their corresponding simulated signal.*

## 2.4.2. Flow properties

Figure 9 shows the effect of different average flow parameter values on the simulated CSF and blood signal. Figure 9a-b show that higher flow velocities result in increased signal attenuation, as expected when using DANTE. However, the velocity at which the signal converges differs between tissue types due to their different relaxation properties and, in particular for blood, $B_1^+$ effects and flow directions. For example, the required velocity to reach a signal of less than 0.02 $M_0$ is around 0.2 cm/s for CSF and 1.6 cm/s for blood.

This velocity threshold is the highest for blood because of the velocity responses of both DANTE and SPACE. During DANTE, the lower $B_1^+$ in the neck means that higher velocities are required for DANTE to be effective. During the SPACE readout, the flow direction of blood is nearly perpendicular to the readout gradient direction. As a result, higher velocities are required to obtain the inherent black-blood SPACE contrast (even without DANTE). Since the typical flow velocity of blood remains substantially higher than the required velocity to achieve black-blood contrast, this should not be a limitation in practice. For CSF, the average absolute velocity of 0.37 cm/s[20] is closer to the critical velocity required to achieve sufficient signal suppression, so increases in CSF signal level can be expected with slight changes in the average velocity or flow direction.

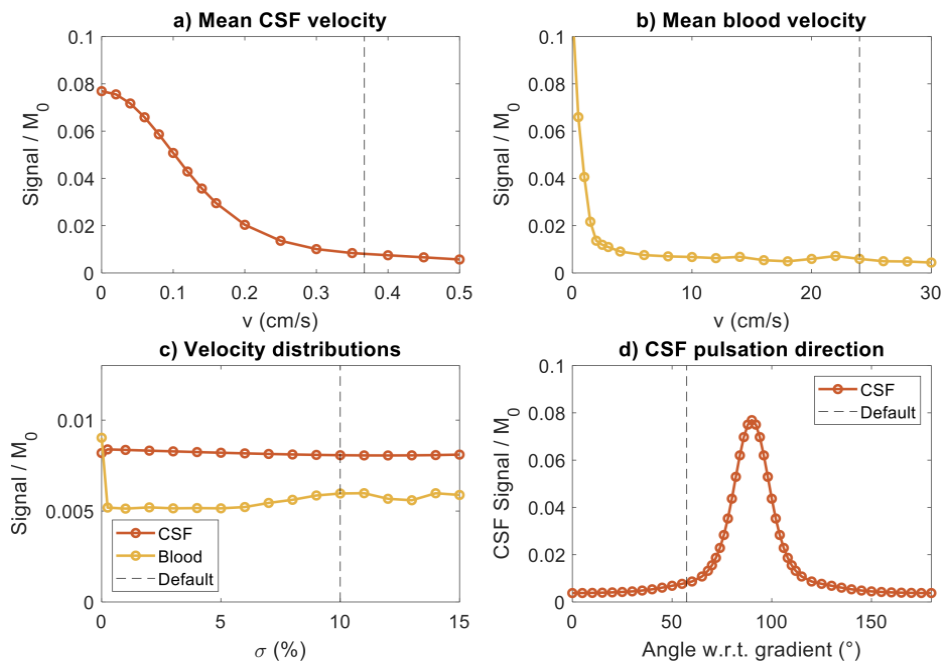

*Figure 9: The effect of the selected flow parameter values on the simulated signal levels of CSF and blood. Results are shown for different (a) mean CSF velocities, (b) mean blood flow velocities, (c) the standard deviation of velocity distributions, and*

*(d)* *CSF pulsation directions vs the DANTE gradient direction. Dashed lines indicate the values generally used in the simulation model.*

The effects of varying the standard deviation in the velocity distribution or the direction of CSF pulsation versus the DANTE gradient direction are shown in Figure 9c-d. Figure 9c indicates that although velocity averaging substantially affects the simulated signal of blood, the exact choice of (non-zero) value for σ does not have a large effect on the resulting averaged signal.

The direction of CSF pulsation (Figure 9d), defined as the angle between the pulsation direction and the DANTE gradient vector, results in substantially reduced CSF suppression for near 90° angles. Since a 90° angle corresponds to having no effective velocity along the gradient direction, agreeing with the simulation for $v_{CSF} = 0$ cm/s shown in Figure 9a.

### 2.5. Accounting for vessel wall motion

Thus far, fully stationary vessel walls have been assumed. However, previously presented *in vivo* acquisitions[17] showed signal variations in the vessel wall, which can be explained using the presented simulation model by including a very slow pulsation of vessel wall tissue. This is shown in Figure 10 for various mean blood pulsation velocities when using two previously proposed DANTE implementations[16,17]. Here, VW pulsation was assumed to follow the same pattern as CSF (Figure 3a) since it includes periods of both positive and negative velocity.

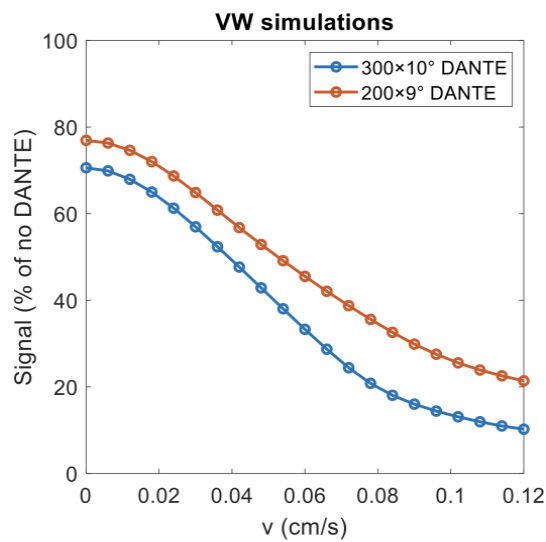

*Figure 10: The simulated signal of VW pulsating at different average velocities in simulations with two different DANTE-preparation settings. Values are expressed relative to the simulation result for stationary VW without DANTE preparation (Equation 2), which corresponds to 0.13 M0.*

To estimate the average velocity of this vessel wall pulsatility such that it can be used in subsequent simulations, we compared *in vivo* acquisitions in 5 healthy volunteers to the VW simulations assuming various mean velocities shown in Figure 10. These preliminary data were acquired on a Siemens (Erlangen, Germany) Magnetom 7T scanner under an approved institutional development ethics protocol using a 1Tx/32Rx head coil from 5 healthy volunteers, using three different T2-weighted DANTE-SPACE protocols:

1. The protocol presented by Viessmann et al.[16], which uses 300 DANTE pulses of 10°;
2. The same protocol but using 200 DANTE pulses of 9°; and
3. The same SPACE protocol but without DANTE preparation.

Using these data, the average VW velocity was estimated such that the resulting signal intensities from simulations agreed with observed (experimental) vessel wall signal changes in hand-drawn VW masks. To quantitatively compare *in vivo* measurements and simulations, both scan data and simulations were expressed relative to the values for stationary VW signal intensity without DANTE preparation. For simulations, this was calculated by expressing the simulated VW signal ($S_{sim}$) as a percentage of the signal from a simulation for stationary VW without DANTE ($S_{NoDante}$):

$$S_{sim}(\%) = \frac{S_{sim}(M_0)}{S_{NoDante}(M_0)} * 100\%. \tag{2}$$

For *in vivo* acquisitions, the 80th percentile of the measured signal intensity values in VW voxels in acquisitions without DANTE ($S_{NoDante,80\%}$) was used as a benchmark for the signal level for stationary VW tissue. From that value, the relative value in each voxel (and for any protocol) was calculated as

$$S_{scan}(\%) = \frac{S_{scan}(SNR)}{S_{NoDante,80\%}(SNR)} * 100\%. \tag{3}$$

The resulting signal distributions were then compared to estimate the average velocity of vessel wall pulsation, which can be used to account for contrast variations due to vessel wall pulsation in simulation-based optimizations.

Figure 10 indicates that for both Protocol 1 and Protocol 2, each simulated signal value corresponds to a single (non-degenerate) mean vessel wall velocity. This means that if *in vivo* vessel wall signal intensity values are also expressed relative to the signal intensity in non-DANTE acquisitions (using Equation 3), it is possible to directly correlate these relative values with the corresponding average velocity in simulations. Based on the preliminary volunteer data acquired using 300 DANTE pulses of 10° (Protocol 1), this suggests an average vessel wall velocity of $0.051 \pm 0.021$ cm/s. The corresponding results based on data acquired from the same volunteers using 200 DANTE pulses of 9° (Protocol 2) suggest an average VW velocity of $0.057 \pm 0.026$ cm/s, in good agreement with the results using Protocol 1. The underlying voxel-wise relative signal and corresponding simulated velocity distributions for both protocols are shown in Supporting Information Figures S4 and S5.

## 3. Discussion

This work presents an EPG-based DANTE-SPACE simulation framework. By including various physiological and spatial variations to enhance the model, it aims to accurately reproduce the contrast mechanisms for the case of *in vivo* acquisitions. This helps explain the mechanisms behind various observed contrasts in acquired data and provides further understanding on how the sequence can be modified to achieve improved contrasts.

### 3.1. Initial validation

The performance of the basic simulation framework design was validated by first separately comparing the performance of the DANTE- and the SPACE-modules relative to literature results, which found good agreement for both. For VW tissue, the addition of DANTE preparation introduces a magnetization decrease in the middle of the SPACE readout. This reduces the VW signal level (as expected when using DANTE[1,16]), but also results in a slight reduction in the VW sharpness. To retain the desired sharpness, future work could account for the magnetization history effects due to DANTE in the calculation of the SPACE variable flip angles.

### 3.2. Accounting for more realistic physiology in the simulations

Accounting for a more realistic physiology affects the simulation results in multiple ways, thereby improving their accuracy. Simulations that account for velocity distributions (Figure 2) increase the temporal stability of the magnetization evolution of moving tissue, resulting in a reduced simulated signal and improved robustness against small changes in the selected mean velocity. Without this velocity averaging, the magnetization over time shows rapidly fluctuating transverse magnetization, similar to what Li et al.[1] showed in DANTE simulations at different specific velocity values.

The introduction of more realistic time-varying pulsatile velocity variation mainly affects the simulations of CSF, which is assumed to oscillate around a central location. This introduces periods of near-zero absolute velocity (which reduce the signal attenuation during DANTE), as well as partial rephasing when the oscillations change direction. In addition to those changes due to pulsation, the inclusion of diffusion reduces the longitudinal magnetization of the CSF at the end of the DANTE-preparation, thereby substantially reducing the resulting CSF signal during SPACE.

The addition of a typical blood flow trajectory only introduces minor changes to the resulting blood signal, as the effective blood flow velocity along the DANTE gradient direction remains sufficiently high despite the different flow directions. However, the differences in $B_1^+$ along the flow trajectory substantially reduce the efficiency of DANTE in low-$B_1^+$ areas. In quadrature (CP) mode at 7T, this increases the measured transverse magnetization during the SPACE readout (Figure 6).

When combined, these simulation enhancements result in a 57% reduction in CSF signal and a 27% reduction in blood signal (for the T2-weighted 7T protocol used by Viessmann et al.[16]) relative to simulations which do not include these enhancements.

### 3.3. Simulations with different tissue properties

The simulations using various T1 and T2 relaxation times (Figure 8) indicate that the (simulated) DANTE-SPACE signal levels are robust to small variations in T1 or T2. Furthermore, Figure 8 shows that the DANTE preparation introduces some T1-weighting. Therefore, when proposing modified protocols, it is important to include simulations for relevant pathologies to ensure that the contrast between healthy and diseased vessel wall tissue remains consistent.

Figure 9 and Figure 10 indicate that for VW and CSF, the respective 'critical' pulsation velocities (where the resulting signal rapidly decreases) are close to their expected velocities. Therefore, small variations in the CSF velocity (Figure 9a) and direction (Figure 9d) can result in substantial reductions in CSF suppression, whereas slow pulsation of the vessel wall also reduces the VW signal. Both of those simulation results are consistent with previous *in vivo* observations for T2-weighted DANTE-SPACE, which showed heterogeneous signal levels in both the VW and the CSF.

### 3.4. Vessel wall motion

Figure 10 indicated that very slow VW motion could result in variable VW signal, which could explain the observed *in vivo* VW signal heterogeneity. This hypothesis was further examined by comparing VW simulations across a range of simulated velocities to the corresponding *in vivo* signal distributions using two different DANTE protocols. This indicates an average VW velocity of $0.051 \pm 0.021$ cm/s based on data with 300 DANTE pulses of 10° (Protocol 1) and $0.057 \pm 0.026$ cm/s based on data with 200 DANTE pulses of 9° (Protocol 2). These values are slightly different, while the underlying vessel dynamics should be the same. This could be explained by the simplifying assumption that all signal difference in the VW is a result of the VW pulsation, which does not account for signal variation due to partial volume effects, sensitivity effects, and different directions of motion relative to the DANTE gradients. Despite this, the resulting values are very similar, and the average value of $0.054$ cm/s can be used as an approximate mean velocity of pulsating vessel wall.

When assuming the VW pulsation to follow the same pattern as CSF, this average velocity of $0.054$ cm/s corresponds to a maximum displacement of around $200$ μm. This result is consistent with displacement simulations in non-stenotic external iliac arteries[30], which have a similar wall thickness to the middle cerebral artery. This again indicates that the value of $0.054$ cm/s provides an adequate approximation. However, this VW motion approximation is based on data from healthy volunteers. A lower VW velocity might be expected in clinical populations due to reduced vessel wall flexibility resulting from pathology and aging[31].

In simulations using the protocol by Viessmann et al.[16], VW moving at 0.054 cm/s is predicted to reduce the resulting VW signal by 46%, leading to a reduction in VW/CSF contrast of 51%. When using simulations to optimize a protocol for achieved contrasts, this introduces a significant new constraint that was not accounted for in previously presented DANTE-SPACE simulation models.

In the work presented here, vessel walls were assumed to pulsate using the same time-varying pattern as CSF (Figure 3a). This pattern was selected since it results in a net-zero displacement during each full cardiac cycle, which is expected for both CSF and VW but not for blood. However, it might be possible to represent the VW pulsation more accurately using other profiles.

### 3.5. Computational considerations

Currently, a single simulation with the computational hardware used in this work takes about 0.16 s for static VW (2 TRs without velocity averaging), 9.6 s for pulsating VW (2 TRs with velocity averaging), 3.6 s for blood (1 TR with velocity averaging), and 7.9 s for CSF (2 TRs with velocity averaging). The computation times of individual simulations could be substantially reduced by parallelizing the individual simulations at each velocity. However, in this work the simulations for different datapoints in parameter sweeps were parallelized instead (with each parallel CPU computing a few full simulations instead of parts of all simulations), to achieve a similar net acceleration.

## 4. Conclusion

An EPG-based DANTE-SPACE simulation framework is presented that includes physiological and spatial variations to accurately reproduce *in vivo* contrast mechanisms. This helps explain observed contrasts in acquired data, such as the VW signal heterogeneity which can be explained by simulations as being due to slowly pulsating vessel walls. Finally, this simulation framework facilitates a more comprehensive optimization of the DANTE-SPACE sequence parameters.

## Acknowledgements

The Wellcome Centre for Integrative Neuroimaging is supported by core funding from the Wellcome Trust (203139/Z/16/Z). We also thank the Dunhill Medical Trust and the NIHR Oxford Biomedical Research Centre for support (PJ). MdB acknowledges studentship support from Siemens Healthineers and the Dunhill Medical Trust. AH acknowledges support from the BHF Centre of Research Excellence, Oxford (RE/13/1/30181).

## Conflict of Interest

PJ is the Editor-in-Chief of Magnetic Resonance in Medicine. In line with COPE guidelines, he recused himself from all involvement in the review process of this paper, which was handled by an Associate Editor. He and the other authors have no access to the identity of the reviewers.

## Data availability statement

In support of Magnetic Resonance in Medicine's reproducible research goal, the MATLAB code for the simulation framework is openly available online at [git.fmrib.ox.ac.uk/ndcn0873/dantespace_epg](git.fmrib.ox.ac.uk/ndcn0873/dantespace_epg). The main simulation tool is `epg_dantespace.m`. The script `example.m` shows examples of how to run the simulations after initializing both the desired sequence parameters (`set_dantespace_parameters.m`) and the tissue properties (`set_tissue_parameters.m`). Finally, a single script that can be used to reproduce all figures in this paper is included (`simulations_paper.m`), as well as the required underlying data (B1$^+$ map and vessel trajectory) and other tools. The simulations require the MATLAB *Image Processing*, *Statistics and Machine Learning*, *Curve Fitting*, and *Parallel Computing* toolboxes.

# References

1. Li L, Miller KL, Jezzard P. DANTE-prepared pulse trains: A novel approach to motion-sensitized and motion-suppressed quantitative magnetic resonance imaging. Magn Reson Med 2012;68:1423–1438 doi: 10.1002/mrm.24142.

2. Li L, Chai JT, Biasiolli L, et al. Black-blood multicontrast imaging of carotid arteries with DANTE-prepared 2D and 3D MR imaging. Radiology 2014;273:560–569 doi: 10.1148/radiol.14131717.

3. Mugler JP. Optimized three-dimensional fast-spin-echo MRI. Journal of Magnetic Resonance Imaging 2014;39:745–767 doi: 10.1002/jmri.24542.

4. Xie Y, Yang Q, Xie G, Pang J, Fan Z, Li D. Improved black-blood imaging using DANTE-SPACE for simultaneous carotid and intracranial vessel wall evaluation. Magn Reson Med 2016;75:2286–2294 doi: 10.1002/mrm.25785.

5. Wan L, Zhang N, Zhang L, et al. Reproducibility of simultaneous imaging of intracranial and extracranial arterial vessel walls using an improved T1-weighted DANTE-SPACE sequence on a 3 T MR system. Magn Reson Imaging 2019;62:152–158 doi: 10.1016/j.mri.2019.04.016.

6. Jia S, Zhang L, Ren L, et al. Joint intracranial and carotid vessel wall imaging in 5 minutes using compressed sensing accelerated DANTE-SPACE. Eur Radiol 2020;30:119–127 doi: 10.1007/s00330-019-06366-7.

7. Sannananja B, Zhu C, Colip CG, et al. Image-Quality Assessment of 3D Intracranial Vessel Wall MRI Using DANTE or DANTE-CAIPI for Blood Suppression and Imaging Acceleration. American Journal of Neuroradiology 2022;43:837–843 doi: 10.3174/ajnr.A7531.

8. Zhang L, Zhang N, Wu J, Liu X, Chung YC. High resolution simultaneous imaging of intracranial and extracranial arterial wall with improved cerebrospinal fluid suppression. Magn Reson Imaging 2017;44:65–71 doi: 10.1016/j.mri.2017.08.004.

9. Cho SJ, Jung SC, Suh CH, Lee J Bin, Kim D. High-resolution magnetic resonance imaging of intracranial vessel walls: Comparison of 3D T1-weighted turbo spin echo with or without DANTE or iMSDE. PLoS One 2019;14:1–10 doi: 10.1371/journal.pone.0220603.

10. Guo Y, Canton G, Chen L, et al. Multi-Planar, Multi-Contrast and Multi-Time Point Analysis Tool (MOCHA) for Intracranial Vessel Wall Characterization. Journal of Magnetic Resonance Imaging 2022;56:944–955 doi: 10.1002/jmri.28087.

11. Kim DJ, Lee HJ, Baik J, Hwang MJ, Miyoshi M, Kang Y. Improved Blood Suppression of Motion-Sensitized Driven Equilibrium in High-Resolution Whole-Brain VesselWall Imaging: Comparison of Contrast-Enhanced 3D T1-Weighted FSE with Motion-Sensitized Driven Equilibrium and Delay Alternating with Nutation for Tail. American Journal of Neuroradiology 2022;43:1713–1718 doi: 10.3174/ajnr.A7678.

12. Wang J, Helle M, Zhou Z, Börnert P, Hatsukami TS, Yuan C. Joint blood and cerebrospinal fluid suppression for intracranial vessel wall MRI. Magn Reson Med 2016;75:831–838 doi: 10.1002/mrm.25667.

13. Zhang L, Zhu Y, Qi Y, et al. T2-Weighted Whole-Brain Intracranial Vessel Wall Imaging at 3 Tesla With Cerebrospinal Fluid Suppression. Front Neurosci 2021;15:1–11 doi: 10.3389/fnins.2021.665076.

14. Coolen BF, Schoormans J, Gilbert G, et al. Double delay alternating with nutation for tailored excitation facilitates banding-free isotropic high-resolution intracranial vessel wall imaging. NMR Biomed 2021;34:1–10 doi: 10.1002/nbm.4567.

15. Cogswell PM, Siero JCW, Lants SK, et al. Variable impact of CSF flow suppression on quantitative 3.0T intracranial vessel wall measurements. Journal of Magnetic Resonance Imaging 2018;48:1120–1128 doi: 10.1002/jmri.26028.

16. Viessmann O, Li L, Benjamin P, Jezzard P. T2-Weighted intracranial vessel wall imaging at 7 Tesla using a DANTE-prepared variable flip angle turbo spin echo readout (DANTE-SPACE). Magn Reson Med 2017;77:655–663 doi: 10.1002/mrm.26152.

17. de Buck M, Hess A, Jezzard P. Optimized DANTE preparation for intracranial DANTE-SPACE vessel wall imaging at 7T. In: Proceedings of the 29th Annual Meeting of ISMRM. Virtual; 2021. p. 4201.

18. de Buck MHS, Kent JL, Hess AT, Jezzard P. Parallel Transmit DANTE-SPACE for improved black-blood signal suppression at 7 Tesla. In: Proceedings of the 31st Annual Meeting of ISMRM. London, UK; 2022. p. 2092.

19. Kong X, Ma J, Westerhold E, et al. Optimization of T1-Weighted DANTE-SPACE for Intracranial Vessel Wall Imaging at 7T. In: Proceedings of the 32nd Annual Meeting of ISMRM. Toronto, Canada; 2023. p. 3299.

20. Linninger AA, Tsakiris C, Zhu DC, et al. Pulsatile cerebrospinal fluid dynamics in the human brain. IEEE Trans Biomed Eng 2005;52:557–565 doi: 10.1109/TBME.2005.844021.

21. Mohammad SA, Osman NM, Ahmed KA. The value of CSF flow studies in the management of CSF disorders in children: a pictorial review. Insights Imaging 2019;10 doi: 10.1186/s13244-019-0686-x.

22. Hennig J. Multiecho imaging sequences with low refocusing flip angles. Journal of Magnetic Resonance 1988;78:397–407 doi: 10.1016/0022-2364(88)90128-X.

23. Weigel M. Extended phase graphs: Dephasing, RF pulses, and echoes - Pure and simple. Journal of Magnetic Resonance Imaging 2015;41:266–295 doi: 10.1002/jmri.24619.

24. Koning W, De Rotte AAJ, Bluemink JJ, et al. MRI of the carotid artery at 7 Tesla: Quantitative comparison with 3 Tesla. Journal of Magnetic Resonance Imaging 2015;41:773–780 doi: 10.1002/jmri.24601.

25. Rane SD, Gore JC. Measurement of T1 of human arterial and venous blood at 7T. Magn Reson Imaging 2013;31:477–479 doi: 10.1016/j.mri.2012.08.008.

26. Rooney WD, Johnson G, Li X, et al. Magnetic field and tissue dependencies of human brain longitudinal 1H2O relaxation in vivo. Magn Reson Med 2007;57:308–318 doi: 10.1002/mrm.21122.

27. Wyss M, Kirchner T, Ringenbach A, Prüssmann K, Henning J. Relaxation Parameter Mapping Adapted for 7T and Validation against Optimized Single Voxel MRS. Proceedings of ISMRM 2013;26:47–48.

28. Augst AD, Barratt DC, Hughes AD, McG Thom SA, Xu XY. Various issues relating to computational fluid dynamics simulations of carotid bifurcation flow based on models reconstructed from three-dimensional ultrasound images. Journal of Engineering in Medicine 2003;217:393–403 doi: 10.1243/095441103770802568.

29. de Buck MHS, Kent JL, Jezzard P, Hess AT. Head-and-neck multichannel B1+ mapping and RF shimming of the carotid arteries using a 7T parallel-transmit head coil. Magn Reson Med 2023.

30. Li J, Pei M, Bo B, et al. Whole-brain mapping of mouse CSF flow via HEAP-METRIC phase-contrast MRI. Magn Reson Med 2022;87:2851–2861 doi: 10.1002/mrm.29179.

31. Uematsu S, Yang A, Preziosi TJ, Kouba R, Toung TJ. Measurement of carotid blood flow in man and its clinical application. Stroke 1983;14:256–266 doi: 10.1161/01.STR.14.2.256.